\documentstyle[preprint,aps]{revtex}
\begin{document}
\draft
\title{Nonperturbative effects in heavy quarkonia}
\author
{C. A. A. Nunes\cite{email} and F. S. Navarra}
\address{Instituto de F\'\i sica, Universidade de S\~ao Paulo  \\
Caixa Postal 20516, 01452-990, S\~ao Paulo, Brasil}
\author{P. Ring}
\address{Physik-Department der Technischen Universit\"at M\"unchen \\
D-8046 Garching, Germany}
\author{M. Schaden}
\address{Department of Physics, New York University, New York, NY 10003, USA}
\date{September 2, 94}

\maketitle
\begin{abstract}
	An effective hamiltonian for heavy quarkonia is derived from QCD by
separating gluonic fields in background and quantum fields and neglecting
anharmonic contributions.  Mesonic states with nonperturbative gluonic
components are constructed.  These states are invariant under gauge changes of
the background fields and form an orthogonal basis.  The effective hamiltonian
is diagonalized in this basis in a systematic $1/m$- and short distance
expansion. For very heavy quarkonia, we obtain an effective potential similar
to the phenomenological funnel potential.  We compare our method to $2^{\rm
nd}$ order perturbation theory in the background fields and demonstrate its
applicability even for the relatively light charmonium system.  The results to
order $1/m$ for pseudoscalar meson masses and wave functions are shown and
compared with those of the Cornell model.
\end{abstract}


\section{Introduction}
\label{intro}

In phenomenological hadron models nonperturbative gluonic effects are accounted
for in a variety of ways. In non-relativistic potential quark
models\cite{quark}, an effective interquark potential is assumed to result from
them and the effective hamiltonian is diagonalized in the Hilbert space of
quarks only. In flux tube models\cite{flux} one does not completely eliminate
the gluons. Their net effect is to generate a color flux tube between quark and
antiquark that binds them. In bag like models\cite{bag} nonperturbative gluonic
effects appear in the guise of the bag constant. They are taken to generate a
vacuum pressure which counterbalances that of the (perturbative) quarks in the
interior of the bag. Gluons in the bag are assumed to be perturbative and
responsible for a hyperfine interaction. All these models are surprisingly
successful in describing the hadronic spectrum. Except for a few exotic states
there seems to be no need for hybrid states, bound states of quarks and gluons.
Nevertheless, even one gluon exchange, leads to intermediate states, where the
(constituent) quarks are not in a color singlet representation and one may
wonder, whether the nonperturbative part of their interaction (for our purposes
everything that is not one gluon exchange) is really as color blind as is
generally assumed. A better understanding of the relation among the various
model parameters and the structure of the nonperturbative ground state would
also clearly be of interest, and there have been attempts to find a relation
between the bag pressure  and  gluon condensation, and the confining potential
in  nonrelativistic models and the string tension obtained in lattice
calculations\cite{string}.

In this paper we study the influence of a nontrivial gluonic ground state on
the {\it structure} of heavy quarkonia. The nearly nonrelativistic nature of
these mesons makes them ideal probes of ground state properties. Their large
mass allows for a systematic expansion of the interaction at small distances
and we gamble that the lowest dimensional (gluonic) condensate suffices for a
rudimentary description of the nonperturbative vacuum structure in this case.
We estimate the matrix elements of the hamiltonian in an approach very similar
to the one used in QCD-sum rules\cite{sumrules} using background fields to
describe nonperturbative gluons and parametrizing their vacuum matrix elements.
In contrast to the approach taken by Voloshin\cite{volos} and
Leutwyler\cite{leut}, we do not treat the nonperturbative part of the
hamiltonian so obtained as small compared to the (perturbative) coulomb
interaction. This is probably only the case for quarkonia heavier than
botonium\cite{leut}.
It has been previously suggested that the inclusion of the finite correlation
of the gauge invariant correlator
\begin{equation}
\label{correlator}
{\cal G}_{\mu\rho , \nu\sigma}^{(1)}(x)=<0\mid \hbox{Tr} \{F_{\mu\rho}(x)S(x,0)
F_{\nu\sigma}(0)S^\dagger(x,0)\}\mid 0>
\end{equation}
will strongly diminish the importance of couplings to the background fields,
allowing again for a perturbative treatment of them\cite{correl,campos}.
In eq.(\ref{correlator})
$$
F_{\mu\rho}=gT^aF^a_{\mu\rho},
$$
$$
S(x,0)=\hbox{P}\exp \bigl(i\int^1_0 dt x^\mu A_\mu(xt)\bigr),
$$
$$
A_\mu = gT^aA^a_\mu,
$$
S is the color transport operator required for gauge invariance; $T^a$ are the
generators  of the gauge group in the fundamental representation.
The correlation length of (\ref{correlator}) should be no larger than that
provided  by the mass of the lowest physical state which contributes. In a
purely ggluonic theory this can at best be that corresponding to the lightest
glueball. According to lattice estimates, the lowest glueball mass is probably
larger than 1500 MeV\cite{chen}, and the correlation length of
(\ref{correlator}) therefore is probably less than 0.13 fm. The value of 0.22
fm extracted by Di Giacomo et al.\cite{giaco} from a simulation on a modestly
sized lattice is probably compatible with this estimate, once errors and finite
size effects are taken into account. If one accepts the arguments used for
instance in
ref.\cite{campos} the nonperturbative effects due to condensates in heavy
quarkonia would be exceedingly small and of little importance. To reproduce
even the first low lying states of the quarkonia spectrum additional
nonperturbative effects, such as a confining potential are required in the
approach of \cite{campos}.

	We wish to point out however, that the correlation function
(\ref{correlator}) is not the relevant one for heavy quarkonia, because it
describes  the propagation
of gluonic modes in the absence of heavy quarks. This correlator
has very little to do with the low energy interaction in heavy quarkonia:
the heavy $\bar q q$ pair is only rarely destroyed to produce gluonium
intermediate states - this is a Zweig rule suppressed process, necessitating
relatively large momentum and thus having a relatively short correlation
length. In heavy quarkonia there almost always is the $\bar q q$ pair around to
screen the color charge of soft effective gluons being exchanged.
We propose that the relevant correlation function is that of a composite color
singlet (scalar) operator such as $\bar \Psi F \Psi $ in a quarkonium state
$\mid M >$, written schematically as
\begin{equation}
\label{correl2}
{\cal G}_M(x-y)=<M\mid \bigl( \bar \Psi F \Psi\bigr) (x)\bigl( \bar \Psi F \Psi
\bigr) (y) \mid M >\,.
\end{equation}

	Essentially two different kinds of physical intermediate states
contribute to this correlation function:
\begin{enumerate}
\item Glueballs: in this cases the correlation functions factorizes into a
decay matrix element of the quarkonium and the correlation function
(\ref{correlator}) - the correlation length of this contribution is essentially
given by the lowest glueball mass.
It is the Zweig rule suppressed process
we mentioned, and we will neglect this contribution for the reasons given
above.
\item Quarkonium states: they only contribute, if the quarkonia have
admixtures, where the quark-antiquark pair is in an octet configuration (with
additional gluons forming an overall color-singlet meson). If only
quark-antiquark (color-)singlet configurations are considered, the contribution
from the excited quarkonia to (\ref{correl2}) would vanish by color invariance.
Since the mass difference of the lowest quarkonium to the first excited
excitation is of the order of 600 MeV, one estimates that the correlation
function (\ref{correl2}) will have a correlation length of about 1/3 fm -- much
larger than one would expect from glueball intermediate states. As explained in
section \ref{effecham} we do approximately include such correlations by
extending the basis Fock-basis for the description of quarkonia by states in
which the heavy
$Q\bar Q$-pair is in an color-octet configuration, with an additional
(soft) color-octet effective gluon to form the singlet.
\end{enumerate}

 We calculate the short distance expansion of the hamiltonian
matrix elements in a gauge invariant basis of color singlet states\cite{basis},
including only terms up to order $r^2$ and $1/m$.  We then  obtain the
eigenstates by (numerical) diagonalization of this hamiltonian matrix and see
how far we can go in this approach. The previous discussion implies
that  a realistic description of heavy quarkonia
within this short distance expansion can only be obtained if
the (background) gauge invariant basis of ref.\cite{basis} is extended to
include states in which the heavy quark-antiquark pair are in a
color octet representation but coupled to a nonperturbative gluonic background
field\cite{volos,volos82} to form an overall color singlet.   Diagonalization
of the hamiltonian matrix leads to a ``color-octet'' component in the
wavefunction of heavy quarkonia that takes account of the possibility of color
exchange between the valence quarks and the background field. This component
becomes rather large for higher excited states and its coupling to the
``singlet'' is the main reason for the distortion of the coulombic spectrum in
this model. In the pseudoscalar meson channel only the few octet states
constructed in section \ref{gaugeinv} couple and a numerical diagonalization
of the resulting system of differential equations is still quite feasible.

	Recently, calculations of heavy quarkonium annihilation rates have
acquired a rigourous theoretical treatment through general factorization
formulae\cite{bodwin} obtained in the context of the effective field theory,
NRQCD\cite{nrqcd}. In the factorization formulae short-distance coefficients,
calculated in perturbation theory, are combined with long-distance matrix
elements that can so far only be computed in lattice simulations. This approach
however provided the solution of the  long-standing problem of infrared
divergences in $h_C$ and $\chi_{c1}$ decays into light hadrons. The probability
for the $\bar q q$ pair to be in a color-octet s-wave at the origin is an
essential input in this approach. Since accurate lattice calculations of this
probability are not yet avaiable, it is determined
phenomenologically\cite{braaten}.
In the approach we propose, this
probability can in principle be computed, since such octet
configurations are included in the description of the meson from the
outset. In this sense, our approach could be considered an inexpensive
alternative to lattice simulations.

In section \ref{effecham} we derive the effective hamiltonian appropriate for
the description of heavy quarkonia. We first obtain the effective Lagrangian to
order $1/m$ by a Foldy-Wouthuysen\cite{foldy} transformation. Using the
background field formalism\cite{back} and neglecting anharmonic quantum
fluctuations one finally arrives at an effective Lagrangian\cite{curci,tese}
that includes background fields up to $2^{\rm nd}$ order and which, in the
instantaneous approximation, gives rise to an effective hamiltonian that is
accurate to order $1/m$ and $r^2$.  Since retardation effects are of order
$1/m^2$, their consistent inclusion would require a much more elaborate
treatment, which would only obscure the basic nonperturbative gluonic effects
we want to elucidate here.

Numerical diagonalization of this effective hamiltonian in the extended basis
then yields (pseudoscalar) meson masses and wave functions.  Our approach
can in principle not be described by an effective potential in the singlet
channel\cite{volos,volos82} because the elimination of the relative octet
states would make it energy dependent.  One can however obtain an energy
independent effective potential for infinitely heavy quarks, where all terms of
order $1/m$ (also the kinetic energy) can be neglected. This potential should
be closely related to the static quark potential one extracts from the Wilson
loop\cite{callan}.  This potential, which we derive and discuss in section
\ref{effecpot}, also gives us an idea how far the short distance
expansion can be
trusted. We show that although it is apparently linearily rising at
intermediate distances $.4fm<r<.7fm$, the potential is also compatible with an
effective exchange of the Gribov type\cite{gribov}.

In section \ref{numres} we compare our method to methods relying on
$2^{\rm nd}$-order perturbation theory in the background field and to the
phenomenological Cornell\cite{cornell} potential. We first present results
where all potential matrix elements of order $1/m$ are neglected.  This greatly
simplifies the calculations because only a few basis states couple, but
pseudoscalar and vector mesons are degenerate at this level. We also give the
results of a more complete calculation of the pseudoscalar quarkonia which
includes the $1/m$ corrections.

Section \ref{concl} is a summary and discussion of our results.

\section{Gauge Invariance and Basis States}
\label{gaugeinv}

It is usually assumed that physical states are color singlets. For heavy
mesons, where the non-relativistic approximation is adequate, one can represent
quark and antiquark fields by 2-component spinors.  Gauge invariance of the
state requires that color is parallel transported from the the quark to the
anti-quark along some path. With a straight path the quark and anti-quark
anti-commutation relations ensure that the basis states are
orthogonal\cite{basis}.  For simplicity we will restrict ourselves in the
following to pseudoscalar mesons. The simplest gauge invariant pseudoscalar
state is of the form

\begin{equation}
\label{singlet}
\mid 2 1 \rangle_S={1\over \sqrt 6}\sum_{ab;\alpha}u^{\dagger a}_\alpha
(\vec x_2) T_{ab}(\vec x_2,\vec x_1 )v^b_\alpha(\vec x_1)\mid \Omega \rangle\,,
\end{equation}
where $u^{\dagger a}(\vec x_2)$ creates a quark with color $a$ and spin
$\alpha$ at $\vec x_2$. $v$ does the same for an antiquark.  We will refer to
such gauge invariant states as singlet states since the quark anti-quark pair
at vanishing separation is in a color singlet representation.

We choose for the color transport operator
\begin{equation}
T_{ab}(\vec x_2,\vec x_1)=Pexp(-ig\int^{\vec x_2,t}_{\vec x_1,t}dx^\mu
A_\mu(x))_{ab}\,,
\end{equation}
the path ordered exponential (denoted by $Pexp$) of gluon operators $A_\mu(x)$
along a straight line from $\vec x_1$ to $\vec x_2$. It is then relatively
straightforward to show that canonical anti-commutation relations for the quark
and anti-quark operators imply that the singlet meson states (\ref{meson}) are
orthogonal\cite{basis}.

We can also construct gauge invariant basis states where the valence quark
anti-quark pair is in a color octet representation at vanishing separation by
coupling them to chromo-electric or -magnetic fields. We shall call these
states octet states for obvious reasons.  Since the chromomagnetic field $\vec
B=\vec B^a\lambda^a/2$ transforms as a pseudovector and the chromoelectric
field $\vec E=\vec E^a\lambda^a/2$ as a vector under rotations and according to
the adjoint representation of the gauge group, we extend the basis for
pseudoscalar mesons by the states
\begin{equation}
\label{octet1}
\mid 2 1 \rangle_B=\sum_{\alpha \beta} {g \over \pi \phi}u^{\dagger
}_\alpha(\vec x_2)\vec \sigma_{\alpha \beta} \cdot \vec B(\vec x_2,t) T(\vec
x_2,\vec x_1)v_\beta (\vec x_1)\mid \Omega \rangle \,,
\end{equation}
\begin{equation}
\label{octet2}
\mid 2 1 \rangle_{E1}=\sum_{\alpha}{\sqrt{3}g \over \pi \phi}  u^{\dagger
}_\alpha (\vec x_2)\vec E(\vec x_2,t) \cdot (\vec x_2-\vec x_1) T(\vec x_2,\vec
x_1)v_\alpha (\vec x_1)\mid \Omega \rangle \,,
\end{equation}
and
\begin{equation}
\label{octet3}
\mid 2 1 \rangle_{E2}=\sum_{\alpha\beta}{i\sqrt{3}g \over \sqrt{2}\pi \phi}
u^{\dagger }_\alpha (\vec x_2)\vec E(\vec x_2,t) \cdot (\vec
\sigma_{\alpha\beta}
\times (\vec x_2-\vec x_1))
T(\vec x_2,\vec x_1)v_\beta (\vec x_1)\mid \Omega \rangle \,,
\end{equation}
the summation over color indices being implied.

The above states are seen to be mutually orthogonal and normalized by the
canonical anti-commutation relations of the quark and anti-quark operators if
we assume expectation values

\begin{eqnarray}
\label{phi}
\langle \Omega &\mid &{g^2\over 4\pi^2} B^{ia}B^{jb} \mid \Omega \rangle=
-\langle \Omega \mid {g^2\over4\pi^2} E^{ia}E^{jb} \mid \Omega \rangle
\nonumber \\
&=&{1\over 96}\delta^{ij} \delta^{ab} \langle \Omega \mid {\alpha\over \pi}
F^{\mu \nu c}F_{\mu \nu}^c \mid \Omega \rangle={1\over 96}\delta^{ij}
\delta^{ab} \phi^2\,,
\end{eqnarray}

and $<E>=<B>=<E B>=0$, which are a consequence of the Lorentz- and parity-
invariance of the vacuum state $\mid\Omega\rangle$. Its nontrivial nature is
reflected in a non-vanishing value for $\phi^2$, which from QCD sumrule
estimates should be close to $(330 MeV)^4$\cite{sumrules}.  If we neglect
expectation values of all higher dimensional gluonic operators the states
(\ref{singlet}),(\ref{octet1}),(\ref{octet2}) and (\ref{octet3}) form a
complete orthogonal basis for the valence quarks of heavy pseudoscalar mesons,
while the singlet states (\ref{singlet}) are only complete in this sense if one
neglects the non-trivial vacuum structure altogether.

We thus effectively truncate the Fock space by taking basis states whose
gluonic sector contains at most 1 soft gluonic mode, i.e. we replace the low
momentum gluonic modes with octet quantum numbers by zero-modes. We show in
section \ref{effecham} that such a severely truncated basis is adequate fo the
problem at hand.

Note also that Lorentz invariance of the ground state $|\Omega>$ forces one to
assume that the chromoelectric background field is either antihermitian or that
the ``E-states'' have negative norm (see (\ref{phi})). Both possibilities lead
to a nonhermitian hamiltonian matrix, whose eigenvalues in general are not
real. Already the construction of basis states for the mesons on a non-trivial
ground state indicates that one can at best hope to find a few stable mesons in
this approach.  We assume that  a pseudoscalar quarkonium $\eta$ is well
described by a linear combination of the above basis states
\begin{equation}
\label{meson}
\mid \eta \rangle = \sum_{M=S,E1,E2,B}\int_{1,2} \psi_M(2,1)\mid 2,1 \rangle_M
\,.
\end{equation}

\section{Effective Hamiltonian}
\label{effecham}

We would like to use the basis constructed in section 2 to approximately
diagonalize the QCD hamiltonian for heavy quarkonia. Since low lying heavy
quarkonia are generally believed to be of relatively small size, the
interaction should be dominated by the perturbative gluon exchange, which leads
to a coulomb-like effective potential. The experimental spectrum deviates
however noticeably from a purely coulombic one and can be reasonably well
reproduced by the combination of a coulomb- and a linear confining- force. We
wish to emphasize here that the truly long range part of the confining force
($r>1{\rm fm}$) is not really tested by the observed quarkonium
states\cite{buch}. Our conjecture is, that what has to be included in a
systematic approach are the first short distance corrections to the
(perturbative) coulomb force due to the non-trivial structure of the gluonic
ground state. This was proposed previously but without any tangible results,
because the corrections were found to be exceedingly large within the framework
of $2^{\rm nd}$-order perturbation theory except for quarkonia beyond
bottonium\cite{volos,leut}. This apparent failure of an idea which we
believe is quite well founded, has led us to reexamine the basic
procedure used in the evaluation of these effects.

In this section we outline the derivation of the effective hamiltonian for
heavy quarkonia, whose matrix elements are correct up to order $\alpha,1/m$ and
$r^2$. To this order in the short distance and heavy quark expansion the
nonperturbative aspects of the ground state can be described by the gluon
condensate $\phi^2$.

The nonrelativistic approximation for heavy quarks is conveniently obtained
from the QCD lagrangian

\begin{equation}
{\cal L}(x)  = -{1 \over4}F^a_{\mu\nu}F^{\mu\nu}_a +\bar \psi
(i\partial\kern-.5em\slash + gT^aV\kern-.5em\slash_a) \psi -m\bar \psi \psi \,,
\end{equation}
by a Foldy-Wouthuysen transformation\cite{foldy}. In terms of transformed quark
fields
\begin{eqnarray}
\psi&\rightarrow&\exp ({i\vec \gamma\cdot \vec D/2m})\psi \,,\nonumber \\
\bar \psi &\rightarrow&\bar \psi \exp({-i\vec \gamma\cdot
D\kern-.5em^{^\leftarrow} /2m})\,,
\end{eqnarray}
the lagrangian is
\begin{eqnarray}
{\cal L}_{NRQCD}(x) & =& -{1 \over4}F^a_{\mu\nu}F^{\mu\nu}_a +\bar \psi
(i\gamma^0D_0-m)\psi \nonumber \\ &&+ \bar \psi ({\gamma^0 \over 2m}[\vec
\gamma \cdot \vec D, D_0]+{(i\vec \gamma \cdot \vec D)^2 \over 2m}))
\psi+O(1/m^2) \nonumber \\ &=&-{1 \over4}F^a_{\mu\nu}F^{\mu\nu}_a +\bar \psi
(i\gamma^0D_0-m)\psi+\bar \psi {\vec D^2 \over 2m} \psi \\ &&- \bar \psi
{ig\vec
\alpha \cdot \vec E \over 2m} \psi +\bar \psi {g\vec \Sigma \cdot \vec B \over
2m} \psi + O(1/m^2) \,, \nonumber
\end{eqnarray}
where $\vec D=\vec \partial -ig\vec V$ and $\vec\Sigma=\left( {\vec\sigma\quad
0\atop 0\quad\vec\sigma}\right)$ does not couple upper and lower spinor
components. They are only coupled in order $1/m$ by the $\vec \alpha$ matrices.
It can be reduced to order $1/m^2$  by another transformation
\begin{equation}
\psi \rightarrow \exp (-ig\vec \alpha \cdot \vec E/ 4m^2)\psi\,,
{\rm etc..}
\end{equation}
The non-relativistic lagrangian to order $1/m$ finally is
\begin{equation}
{\cal L}_{NRQCD}(x) =-{1 \over4}F^a_{\mu\nu}F^{\mu\nu}_a +\bar \psi
(i\gamma^0D_0-m)\psi+\bar \psi {\vec D^2 \over 2m} \psi +\bar \psi {g\vec
\Sigma \cdot \vec B \over 2m} \psi\,,
\end{equation}
which can be written in terms of uncoupled 2-component spinors by decomposing
$\psi=(u,v^\dagger)\,\ \bar\psi=(u^\dagger,-v)$ and using the Dirac
representation of the $\gamma$-matrices.

To effect a short distance expansion we separate the gluonic fields in slowly
varying background-\cite{back}\-($A$) and quantum- ($Q$) fields ha\-ving high
fou\-rier com\-po\-nents:
\begin{equation}
\label{separ}
V^a_\nu = A^a_\nu+Q^a_\nu \,.
\end{equation}

The division (\ref{separ}) can however only be defined if the gauge is fixed.
In deriving the spin-dependence of nonperturbative interactions for heavy
quarkonia Curci et al.\cite{curci} found a particular gauge very convenient. We
essentially follow their procedure here and impose the Coulomb background gauge
condition
\begin{equation}
\label{quantum}
D_iQ^i=0
\end{equation}
for the quantum fields, where
\begin{equation}
D_\mu Q_\nu = \partial_\mu Q_\nu + gA_\mu
\times Q_\nu
=\partial_\mu Q_\nu +gf^{abc}A_{\mu b} Q_{\nu c} \,.
\end{equation}
The background fields are defined in a modified Schwinger gauge\cite{curci}
\begin{equation}
\label{gauge}
A^b_j= -{1 \over2} F^b_{ji} x^i \qquad ; \qquad A ^b_0= - F^b_{0i} x^i\,,
\end{equation}
valid to order $x^2$, where we assumed that the field-strengths corresponding
to the background  fields are constant (or have sufficiently low momenta, such
that they  can be regarded  as  essentially constant over the extent of the
meson). This definition of the background fields in terms of (practically)
constant field strengths also gives a definite meaning to the separation in
equation (\ref{separ}). It also implies that the background fields in this
gauge are (practically) time independent -- a property which will become useful
when a Hamiltonian is required.

We next expand the nonrelativistic Lagrangian only to second order in the
quantum fields and subsequently integrate them out in favor of an effective
(coulombic) interaction. Although ghost terms are necessary in the gauge
defined by (\ref{quantum}), they do not contribute to quadratic order in the
quantum fields. Our truncation of the interaction terms for the quantum fields
eliminates all radiative corrections. To obtain them one would have to go
beyond this approximation and calculate perturbative corrections before
eliminating hard gluons.  Fortunately, the asymptotic freedom of QCD guarantees
that they are only logarithmic at short distances and could be accounted for by
a running coupling constant. These logarithmic corrections to the Coulomb
potential do not seem to be dramatically important for describing heavy
quarkonia spectra\cite{quigg} and we will not include them in this study.

The matrix elements of the slowly varying background fields ($A$) will however
be parameterized. Their amplitude is large and an expansion in the coupling in
this case not applicable.

After elimination of the quantum gluonic fields by their equations of motion
the effective lagrangian in terms of the background and heavy quark fields
becomes\cite{tese}:

\begin{eqnarray}
\label{lagr}
L_{eff} &=&  \int d^4x\{ -{1\over 4}F^{\mu\nu}F_{\mu\nu}+
\bar \psi (i\gamma^0\partial_0 +
g\gamma^0 A_0-m ) \psi  \nonumber \\ &&+{1\over 2m} \bar \psi (\vec
\bigtriangledown^2-ig\vec \bigtriangledown \cdot
\vec A -ig\vec A \cdot \vec \bigtriangledown -g^2\vec A^2)\psi
+{g\over 2m} \bar \psi \vec \Sigma \cdot \vec B \psi \nonumber \\ &&+ g^2\int
d^4y  \psi^\dagger (x) T^a \psi (x) {\cal D}^{ab}(x,y)
\psi^\dagger (y) T^b \psi (y)  \\
&&- {1 \over 2}\int d^4y  J^a_i(x)(\tilde {\cal D}^{ab}_{ij}(x,y)-{\cal
K}^{ab}_{ij} (x,y))J^b_j(y) \,, \nonumber
\end{eqnarray}
where
\begin{eqnarray}
J^a_k(x)&=&-{ig\over 2m}[(\partial_k\bar \psi )T^a\psi - \bar \psi
T^a\partial_k \psi -i\epsilon_{ijk}\partial_i (\bar \psi T^a \Sigma_j \psi)]
\nonumber \\
&&+ {g^2 \over 2m} \bar \psi (T^bT^a+T^aT^b)\psi A^a_k -{g^2 \over
2m}f^{abc}\epsilon_{ijk}A^b_i\bar \psi T^c\Sigma_j \psi
\nonumber \\
&&-2g^2 f^{adc} F^d_{k0}\int d^4z {\cal D}^{cb}(x,z)\psi^\dagger (z)T^b\psi(z)
\,.
\end{eqnarray}
       The propagator ${\cal D}$ relates the $Q_0$ field to its source
\begin{equation}
Q^a_0(x)=\int d^4y {\cal D}^{ab}(x,y)j^0_b(y)\,,
\end{equation}
with
\begin{equation}
j^a_0= g \psi^\dagger T^a\psi + 2gf^{abc}F_{bi0}Q_{ci}\,,
\end{equation}
and satisfies
\begin{equation}
(D_jD_j)^{ab} {\cal D}^{bd}(x,y)=\delta^{ad}\delta^4(x-y)\,.
\end{equation}

       Similarly the propagator $\tilde {\cal D}$ appears in the elimination of
the spatial components $Q_i$ and satisfies
\begin{equation}
\int d^4z M^{ab}_{ij}(x,z) \tilde {\cal D}^{bc}_{jk}(z,y) = \delta^{ac}\delta^4
(x-y)\delta_{ik}\,,
\end{equation}
   with the differential operator $M^{ab}_{ij}$ given by
\begin{eqnarray}
M^{ab}_{ij}(x,z)&=&\delta^4(x-z)[-(D_\mu D_\mu)^{ab}\delta_{ij}- 2g f^{abc}
F^c_{ij}\nonumber \\ &&-{g^2 \over 2m}\bar \psi(T^aT^b+T^bT^a)\psi \delta_{ij}+
{g^2\over 2m}f^{abc} \epsilon_{ijk} \bar \psi T^c\Sigma^k \psi] \\ &&+4g^2
f^{dfa} f^{ceb} F^f_{i0}(x) {\cal D}^{dc}(x,z)F^e_{j0}(z)\,.
\end{eqnarray}
   Finally, the propagator ${\cal K}$ enters when the lagrange multiplier of
the gauge fixing condition for the quantum fields (\ref{quantum}) is eliminated
in turn. Its equation of motion is
\begin{equation}
\int d^4z D^{ba}_i(x) \tilde {\cal D}^{ad}_{ij} (x,z) D^{dc
}_j(z) {\cal K}^{cd}(z,y)=\delta^{bd} \delta (x-y)\,.
\end{equation}

   These rather formidable integro-differential equations for the Green
functions can formally be solved order by order in the background field $A$.
Since we will only retain terms of the hamiltonian matrix proportional to the
lowest dimensional condensate $<g^2FF>$, we only keep terms up to second order
in the background fields in this gauge (\ref{gauge}). As we will see shortly,
linear terms in the background field have to be retained, although we will
assume that $<F>=0$, e.g. that global colour- and lorentz- invariance is not
broken.

In order to obtain a tractable hamiltonian, further approximations are
necessary. We will neglect all retardation effects  in the effective
interaction.  This instantaneous approximation is correct to order $1/m$,
because retardation effects are generally expected to be of order $1/m^2$.
Since the interaction is instantaneous, the effective hamiltonian becomes time
independent even in the presence of the background fields (which are (nearly)
time independent in the gauge (\ref{gauge})). This greatly simplifies the
interpretation of our results.

The effective hamiltonian, correct to order $1/m$, $r^2$ and $\alpha_s$
therefore is
\begin{eqnarray}\label{hamilt}
H&=&\int d^3x \Biggl\{ u^\dagger (\vec x)mu(\vec x)+v(\vec x)mv^\dagger (\vec
x) -\frac{1}{4} F_{\mu\nu}F^{\mu\nu} \nonumber \\ &&\ -u^\dagger (\vec
x)T^AgE^A_ix_i u(\vec x) - v(\vec x)\bar T^A gE^A_ix_i v^\dagger (\vec x)
\nonumber \\ &&\ -u^\dagger (\vec x)\frac{\vec \bigtriangledown^2_x}{2m} u(\vec
x) -v(\vec x) \frac{\vec
\bigtriangledown^2_x}{2m} v^\dagger (\vec x)\nonumber \\
&&\ +\alpha\int d^3y u^\dagger(\vec x)T^Au(\vec x)\frac{1}{r}v(\vec y)\bar
T^Av^\dagger(\vec y) \\ &&\ +\frac{\alpha}{2}\epsilon_{ijk}gB^C_kf^{ACB}\int
d^3y u^\dagger(\vec x)T^Au(\vec x)\frac{y_jx_i} {r}v(\vec y)\bar
T^Av^\dagger(\vec y) \nonumber \\ &&\ -\frac{\pi^2 \phi^2\alpha}{64}\int d^3y
u^\dagger(\vec x)T^Au(\vec x)\frac{(\vec x \times \vec y)^2} {r}v(\vec y)\bar
T^Av^\dagger(\vec y) \nonumber \\ &&+\frac{1}{m}\Biggl[\frac{\pi^2
\phi^2}{128}u^\dagger (\vec x)u(\vec x)
\vec x^2
+\frac{\pi^2 \phi^2}{128}v(\vec x)v^\dagger (\vec x)\vec x^2 \nonumber \\
&&\quad -\frac{i}{2}u^\dagger (\vec
x)T^A\epsilon_{ijk}gB^A_kx_j\partial_{x_i}u(
\vec x)
-\frac{i}{2}v(\vec x)\bar T^A\epsilon_{ijk}gB^A_kx_j\partial_{x_i}v^\dagger (
\vec x)
\nonumber \\
&&\quad -\frac{1}{2}u^\dagger (\vec x) \sigma_iT^A gB^A_iu(\vec x)-
\frac{1}{2}v(\vec x) \sigma_i\bar T^A gB^A_iv^\dagger (\vec x)  \nonumber \\
&&\quad +\frac{i\alpha}{8}f^{ADC}gE^D_j
\Bigl[(\partial_{x_i}u^\dagger(\vec x))
T^A u(\vec x) -u^\dagger(\vec x)T^A(\partial_{x_i}u(\vec x))
\nonumber \\
&&\quad \quad -i\epsilon_{ilk}\partial_{x_l}(u^\dagger(\vec x)T^A\sigma^ku(\vec
x)\Bigr]
\cdot
\int d^3y(-3\delta_{ij}r+\frac{r_ir_j}{r})v(\vec y)\bar T^Cv^\dagger (\vec y)
\nonumber \\
&&\quad -\frac{i\alpha}{8}f^{ADC}gE^D_j \Bigl[-(\partial_{x_i}v(\vec x))
\bar T^A v^\dagger (\vec x) +v(\vec x)\bar T^A(\partial_{x_i}v^\dagger
(\vec x)) \nonumber \\ &&\quad \quad +i\epsilon_{ilk}\partial_{x_l}(v(\vec
x)\bar T^A\sigma^kv^\dagger (\vec x)
\Bigr]
\cdot
\int d^3y (-3\delta_{ij}r+\frac{r_ir_j}{r})u^\dagger (\vec y)T^Cu(\vec y)
\Biggr]
\Biggr\}\,.
\nonumber
\end{eqnarray}
Here $u(\vec x)$ and $v(\vec x)$ denote the anihilation operators for a quark
and antiquark of mass $m$ respectively whose spin and color indices have been
suppressed, $\vec r=\vec x -\vec y$ and $T^A,\bar T^A$ are the hermitian
generators of the SU(3) color Lie-algebra in the $3$ and $\bar 3$
representation.

This is essentially the Hamiltonian to order $1/m$ derived previously by Curci
et al.\cite{curci} in position space, except for a term which can be regarded
as a long range correction of the Coulomb potential
\begin{equation}\label{coulcor}
\frac{3\pi^2 \phi^2 \alpha}{64}\int d^3x d^3y u^\dagger(\vec x)T^A u(\vec x)\,
r^3\,v(\vec y)\bar T^A v^\dagger(\vec y)\,.
\end{equation}
We have disregarded this term because it is of order $r^3$ and our gauge fixing
condition (\ref{gauge}) is not valid at this order.

The terms of the Hamiltonian (\ref{hamilt}) linear in the chromo -electric and
-magnetic fields as well as those proportional to $\vec x^2$ are obviously also
not translationally invariant and therefore apparently depend on the chosen
gau\-ge fi\-xing point in (\ref{gauge}).  This gauge dependence of
(\ref{hamilt}) is however absent\cite{tese} in its matrix elements in the basis
(\ref{singlet}),(\ref{octet1}),(\ref{octet2}) and (\ref{octet3}). It cancels
against terms which appear when derivatives act on the color transport matrix
of the states.  This cancelation of course only works to a certain order in the
short distance expansion and only occurs if the Hamiltonian and the basis in
which it is diagonalized are defined consistently. This is not a big surprise
in any gauge theory, but a gauge invariant scheme to include nonperturbative
aspects of the ground state in a hamiltonian formulation without losing gauge
invariance was only first proposed in \cite{basis}, but not used to its full
extent there. From the above we see that a short distance expansion in the
construction of the basis as well as the hamiltonian to order $r^2$ can be
performed and the effects from a nontrivial expectation value $\phi^2$
consistently included.

The straightforward but lengthy computation of the matrix elements of the
Hamiltonian (\ref{hamilt}) in the basis for pseudoscalar mesons (\ref{meson})
to obtain the coupled differential equations (\ref{diffeq}) for the
wavefunctions\cite{tese} will not be exhibited here. We do however also have to
account for matrix elements of the purely gluonic part of the Hamiltonian
\begin{equation}
\label{gluonic}
H_G=\int d^3 x\,{\cal{H}}_G(x)\,,\qquad {\cal{H}}_G(x)=(E^2+B^2)/2 + O(\alpha)
\end{equation}
in the basis states (\ref{singlet}),(\ref{octet1}),(\ref{octet2}) and
(\ref{octet3}). The order $\alpha$ terms in (\ref{gluonic}) are due to
the 1-loop corrections from the quantum fields. When these
corrections are properly included, the above Hamiltonian should be consistent
with the trace anomaly\cite{trace} and reproduce
the relation between the energy density of the nonperturbative gluonic ground
state and the condensate value.
All we
will need in the sequel is that this
correction is a local operator of dimension 4 and order $\alpha$
(such as $(11\alpha/16\pi) (E^2-B^2)$).

{}From rotational symmetry we conclude that only diagonal matrix
elements of $H_G$ can be
non-vanishing. Since we neglect expectation
values of operators with dimension greater than 4, matrix elements in
``octet''-states
(\ref{octet1}),(\ref{octet2}),(\ref{octet3}) vanish in this approximation.

{}From (\ref{phi}) we might naively infer that the matrix elements of $H_G$
between ``singlet'' states vanish as well.
Phenomenologically we do however need an energy splitting between the
``singlet''- and ``octet''- states. Leutwyler\cite{leut2} attributes it to an
effective mass of the low frequency gluonic modes.
A more careful examination of contributions from $H_G$ reveals the origin of
such an effective mass in the present approach. To obtain the energy of the
meson relative to that of the vacuum, one has to commute the purely gluonic
hamiltonian with the creation operator for the meson.
As discussed previously, gauge invariance requires that this creation operator
contains a gluonic string, or color transport matrix. Within our
approximation, we therefore have to consider the commutator of the
creation operator for the singlet component with the local
Hamiltonian density of ${\cal{H}}_G(x)$.
Defining the creation operator for the singlet component
of the meson by $$\int_{1,2}\psi_S(1,2)\mid
2,1\rangle_S=O^+_S\mid\Omega\rangle\,,$$
the contribution of the lowest dimensional operator
to the commutator is of the form
$$[{\cal H}_G(x), O^+_S]= O^+_S A(x) \theta_V(x)\,, $$
where $A(x)$ is a gauge invariant scalar operator of dimension 4 and
order $\alpha$ (because the $g^0$-part of the string obviously
commutes) and $\theta_V(x)$ is a dimensionless
c-number function that vanishes for $x$ outside the
meson (because the string in $O^+_S$ and $x$ are otherwise separated by
a spacelike distance). Since we do not neglect the vacuum expectation of the
gauge invariant operator $\alpha F^2$ in our approach, the matrix
element in the singulet channel within our approximation becomes
\begin{equation}
\label{gluen}
\int d^3 x\,_S\langle 2,1\mid [{\cal H}_G(x), O^+_S]\mid
\Omega\rangle=C\Psi_S(1,2)\,,
\end{equation}
where $C$ is a constant which depends on the exact nature of the
function $\theta_V(x)$.

Although we cannot determine the value of this
constant on theoretical grounds, the above argument shows that it
would
be inconsistent with our approximations to neglect this contribution
of $H_G$ in the singlet channel.A non-vanishing constant $C$ implies that the
lowest order energies of the singulet and octet
states differ by gluonic contributions. From the correlation length of
(\ref{correl2}) discussed in the introduction one would
expect a splitting of the octet and singulet channels of the order of
600~MeV, if the Fock-basis is a reasonable one. Our best fit to the
quarkonium spectrum was obtained with $C=-756$. That this splitting of
the ``singlet'' and ``octet'' Fock states is close to the correlation
length of (\ref{correl2}) indicates that the Fock-space is quite adequate
for this calculation (i.e. the interaction energies are small compared to
the splitting of the bare Fock states).

$m_g=-C$ can  also be interpreted
as an effective mass for the low frequency gluon mode in the
octet-channel (since a small change in
the overall normalization of the energy can be absorbed in a slight
change of the heavy quark mass). This is the  point of view taken by
Leutwyler\cite{leut2}. Our estimate $m_g=-C\sim 756 MeV$ is not in
conflict with the fact that the
lightest glueball has at least two such excitations and a mass of $\sim
1700{\rm MeV}$\cite{chen}, although a somewhat smaller glueball mass
would seem more natural.

Including this energy shift between ``singlet'' and ``octet'' components,
the coupled set of differential equations for flavor singlet
pseudoscalar mesons becomes
\begin{eqnarray}
\label{diffeq}
\Bigl[ 2 m-E+C-\frac{1}{m}\frac{\partial^2}{\partial
r^2}-\frac{4}{3}\frac{\alpha_s}{r}+\frac{\pi^2\phi^2 r^2}{36m} \Bigr] S(r)&=&
\nonumber \\ +(-\frac{\pi \phi
r}{3\sqrt{2}}+\frac{g^2\phi}{4\sqrt{2}m}+\frac{g^2\phi r}{ 8\sqrt{2}
m}\frac{\partial}{\partial r}&)& E_1(r)-\frac{5g^2 \phi}{64m}E_2(r)
\nonumber \\
\Bigl[ 2 m-E-\frac{1}{m}(\frac{\partial^2}{\partial
r^2}-\frac{2}{r^2})+\frac{1}{6}\frac{\alpha_s}{r} \Bigr] E_1(r)=(\frac{\pi
\phi r}{3\sqrt{2}}&+&\frac{g^2\phi}{8\sqrt{2}m}-\frac{g^2\phi r}{
8\sqrt{2} m}\frac{\partial}{\partial r}) S(r)  \nonumber \\
\Bigl[ 2 m-E-\frac{1}{m}(\frac{\partial^2}{\partial
r^2}-\frac{2}{r^2})+\frac{1}{6}\frac{\alpha_s}{r} \Bigr] E_2(r)&=&\frac{5 g^2
\phi}{64m}S(r)\,,
\end{eqnarray}
where $E$ is the mass eigenvalue of the quarkonium and $m$ is the mass of the
constituent quarks.The functions $S(r)$, $E_1(r)$  and $E_2(r)$ are related to
the wave function components in the expansion (\ref{meson}) via \hfill \break
$\psi_S(r)=\frac{1}{r}S(r) \quad ;
\quad \psi_{E_1}(r)=\frac{1}{r^2}E_1(r)\quad ;
\quad \psi_{E_2}(r)=\frac{1}{r^2}E_2(r)$.\hfill \break
 Note that there is no coupling to the pseudoscalar B-states, because it is
proportional to the total spin in (\ref{hamilt}), which vanishes for
pseudoscalar mesons.  As indicated above, these equations only depend on the
relative coordinate $r$ and all reference to a gauge fixing point--present in
(\ref{hamilt})--has disappeared in the evaluation of the effective Hamiltonian
in the gauge invariant basis.

Apart from the constituent quark mass, the only parameters that enter equations
(\ref{diffeq}) are the (reasonably well known) condensate $\phi^2$ and the
constant $C$, whose value we have tried to estimate above, as well as the
strong coupling constant $\alpha_s$.

\section{``Effective Potential''}
\label{effecpot}

Before solving the coupled set of equations (\ref{diffeq}) by numerical
methods, it is instructive to consider the limit of infinitely heavy quarkonia.
In this case, all terms proportional to $1/m$ (including the kinetic energy) in
(\ref{diffeq}) can be dropped, and the resulting equations give the binding
energy $V=E-2 m$ for states where the quark and anti-quark are localized a
distance $r$ apart (i.e. for wave-functions $S(r),E1(r)$ and $E2(r)$ all
proportional to $\delta(r-r_0)$. It is natural to compare this mass-independent
binding energy to phenomenological potentials and those extracted from the
expectation values of Wilson loops in numerical lattice simulations.

In this limit the $E2$-component decouples and one has to solve the algebraic
equations
\begin{eqnarray}
(-V+C-\frac{4}{3}\frac{\alpha_s}{r})S(r)&=&-\frac{\pi \phi r}{3\sqrt{2}}
E_1(r)\nonumber\\ (-V+\frac{1}{6}\frac{\alpha_s}{r})E_1(r)&=&\frac{\pi \phi
r}{3\sqrt{2}} S(r)\,.
\end{eqnarray}
The eigenfunctions are obviously localized and the eigenvalue or effective
potential, $V(r)$, given by
\begin{equation}
\label{pot}
V(r)=-\frac{1}{2}(\frac{7}{6}\frac{\alpha_s}{r}+\sqrt{\frac{9}{4}\frac{
\alpha_s^2}{ r^2}-\frac{9}{3}\frac{\alpha_s C}{r} - \frac{2\pi^2\phi^2r^2}{9} +
C^2})
\end{equation}

Figure \ref{fig1}(a) shows this effective potential and its Coulomb part for
parameters $\phi^2=(360 MeV)^4$, $C=-756$ and $\alpha_s=0.39$, which we found
appropriate for charmonium.

Although this potential is certainly no longer valid for $r>0.9{\rm fm}$, where
the root in (\ref{pot}) becomes purely imaginary it does show a nearly linear
behaviour for intermediate distances $0.4{\rm fm} <r<0.8 {\rm fm}$, with a
correspondingly constant force of $\sim 840{\rm MeV}/{\rm fm}$, which compares
favorably with a string tension of about $800-1000{\rm MeV}/{\rm fm}$ extracted
from recent lattice calculations of the potential\cite{schil}.  In Fig.
\ref{fig1}(b) we compare our effective potential (\ref{pot}) to that extracted
from lattice data\cite{schil} and to the phenomenological potential used by
ref\cite{cornell}.  It is perhaps also of some theoretical interest, that this
potential closely resembles that derived from an ``instantaneous'' Gribov type
gluon exchange\cite{gribov}
\begin{equation}
\label{gribovpot}
V_{Gribov}(r):=-{4\over 3}g^2\int {d^3 k\over (2\pi)^3}{k^2\over k^4+\kappa^4}
e^{i\vec k\vec r}\, =\,-{4\over 3}{\alpha_s\over r} e^{\kappa r/\sqrt{2}}
cos({\kappa r/\sqrt{2}})\,,
\end{equation}
in the limited range of interest $r<.9{\rm fm}$ with $\kappa=500{\rm MeV}$. For
comparison we also show the potential (\ref{gribovpot}) in Fig \ref{fig1}(b).

It is encouraging that our rather crude approximations to the vacuum structure
seem to qualitatively reproduce the potential for very heavy quarks at small
distances. The analytical results of this section justify the numerical
calculation of ``octet''-components in heavy quarkonia which we now present.

\section{Numerical Results}
\label{numres}

Going beyond the static approximation by including the kinetic energy of the
quarks but still neglecting coupling terms of order $1/m$ in (\ref{diffeq}),
the quarkonium spectrum becomes discrete and the following coupled set of
differential equations must be solved numerically
\begin{eqnarray}
\label{leutapp}
\Bigl[ 2 m-E+C-\frac{1}{m}\frac{\partial^2}{\partial
r^2}-\frac{4}{3}\frac{\alpha_s}{r} \Bigr] S(r)&=&-\frac{\pi \phi r}{3\sqrt{2}}
E_1(r) \nonumber \\
\Bigl[ 2 m-E-\frac{1}{m}(\frac{\partial^2}{\partial
r^2}-\frac{2}{r^2})+\frac{1}{6}\frac{\alpha_s}{r} \Bigr] E_1(r)&=&\frac{\pi
\phi r}{3\sqrt{2}} S(r)\,.
\end{eqnarray}
This is essentially Leutwyler's\cite{leut} approximation to the problem, who
obtained the perturbation of the Coulomb spectrum in $2^{\rm nd}$-order of
$\phi$ and found that it is exceedingly large for canonical values of the
condensate. Note that the Coulomb force is repulsive in the ``octet'' channel
and $1/8$-th in strength compared to the ``singlet'' channel (this is just the
ratio of $T^a_1\bar T^a_2$ in the two representations) and the $E_2$-state
still decouples in this approximation. In this approximation, vector- and
pseudoscalar- quarkonia are furthermore still degenerate. This is expected,
since the spin splitting is of order $1/m$. It however is another nontrivial
consistency check of our method, because the basis states for vector mesons are
of course quite different.  Nevertheless equivalent equations to
(\ref{leutapp}) result, if $1/m$ potential terms are neglected.

The relative minus sign of the two off-diagonal coupling terms in
(\ref{leutapp}) shows that this Hamiltonian is not hermitian and that its
eigenvalues will generally not be real. This effect is completely missed if the
coupling is treated as a perturbation. To any finite order in perturbation
theory the correction to the Coulomb spectrum is real. Perturbation theory does
however show that a few (low lying) eigenvalues of (\ref{leutapp}) are real for
a sufficiently small $\phi$. For the canonical value $\phi^2=(300-380{\rm
MeV})^4$ we find numerically $2-4$ stable bound quarkonia in the pseudoscalar
channel, depending on the heavy quark mass $m$.

But even for these low states, the deviation of the exact correction to the
coulomb eigenvalue from the $2^{\rm nd}$-order estimate is large for the
charmonium and bottonium systems as shown in table \ref{tab1}. We conclude with
Leutwyler\cite{leut} that perturbation theory is not applicable for canonical
values of $\phi$, but that the effective non-hermitian Hamiltonian equations
(\ref{leutapp}) do yield reasonable corrections in an exact solution.

The value $C=-756{\rm MeV}$ used by us was ajusted to reproduce the correct
splitting between $\eta_c$ and $\eta_c'$\cite{data} when $1/m$ terms are
included (see below). The corrections to the coulomb spectrum are however not
extremely sensitive to the value of $C$ once it is large enough.  Using $C=
-1400{\rm MeV}$ instead would only reduce the nonperturbative contributions for
the ground and first excited bottonium-states in table \ref{tab1} to $6$ and
$75{\rm MeV}$ respectively.

The fact that a perturbative evaluation in $\phi$ is not appropriate in
(\ref{leutapp}) can also not be circumvented by including loop corrections
(higher orders in $\alpha_s$ than we have treated so far) to the perturbative
coulomb potential. Titard and Yndur\'ain\cite{ynd} recently proposed to modify
the perturbative part of the interaction in the following manner
\begin{equation}
\label{coulcorr}
{\alpha_s \over r}\rightarrow { \alpha_s(\mu^2)\left[ 1+(a_1+\gamma_E
\beta_0/2)\alpha_s(\mu^2)/\pi \right] \over r}+ {\beta_0\alpha_s^2(\mu^2) \over
2\pi}{\log r\mu \over r} \,,
\end{equation}
where the appropriate constants for the SU(3) color group with 4 light quark
flavors are $\beta_0=8.33$ and $\quad a_1=1.47$. The first term of
(\ref{coulcorr}) which contains a piece of one-loop radiative corrections was
taken by Titard and Yndur\'ain as an effective Coulomb potential and solved
exactly. The effective coupling constant is defined as
\begin{equation}
\tilde \alpha_s(\mu^2)= [1+(a_1+\gamma_E \beta_0/2){\alpha_s(\mu^2) \over \pi}]
\alpha_s(\mu^2)\,.
\end{equation}
The second term in (\ref{coulcorr}) was treated to first order in perturbation
theory. A new scale parameter $\mu$ was introduced which depends on the
quarkonium system under consideration. Taking the effective Coulomb potential
we obtain the deviation of the eigenvalues from $2^{\rm nd}$-order perturbation
theory in $\phi$ shown in table
\ref{tab2} for two sets of parameters in the bottonium system (still neglecting
$1/m$ corrections).

The deviation of the exact correction from the perturbative estimate is reduced
somewhat, especially in the second case, but still far from negligible. In
assessing the quality of a perturbative evaluation in this case, one should
also keep in mind that the scale parameter $\mu$ of Ref.\cite{ynd} was chosen
in such a way that the $2^{\rm nd}$ order correction in $\phi$ is precisely
canceled by the correction terms to the effective Coulomb potential in
(\ref{coulcorr}). For the first set of parameters this cancellation occurs for
the ground state energy. The second set was chosen so that the splitting
between the first excited state and the ground state is not affected to $2^{\rm
nd}$ order.  This is obviously a quite arbitrary procedure which requires an
additional parameter and furthermore does not cure the problem that
perturbation theory simply does not apply (as table \ref{tab2} clearly
indicates).

We therefore will not include these modifications to the Coulomb force in our
discussion of the numerical solution to the full set of coupled equations
(\ref{diffeq}). The inclusion of  $1/m$ potential terms lifts the degeneracy of
pseudoscalar- and vector- quarkonia and we restrict our discussion here to the
pseudoscalar case.

The $E_2$-states now couple in, but generally have small (negative) norms,
because the coupling is of order $1/m$. We nevertheless solved the full set of
coupled equations numerically, although a perturbative treatment of the
$E_2$-state admixture would probably have been sufficient. The eigenvalues we
obtained are summarized in table
\ref{tab3} and compared to those obtained by Eichten et al.\cite{cornell} with
the phenomenological funnel potential. Note that we only found 3 or 4 stable
solutions in the charmonium and bottonium systems respectively.  We used the
same values for the quark masses and the coupling constant as Eichten et
al.\cite{cornell}.The constant $C$ was chosen to reproduce the experimental
splitting (not confirmed\cite{data}) between $\eta_c$ and $\eta_c'$. It was not
adjusted in the bottonium system. The gluon condensate value $\phi^2=(360{\rm
MeV})^4$ we used is within QCD-sumrule estimates\cite{reind} for this
nonperturbative quantity. All eigenvalues were finally shifted by $E_0=98{\rm
MeV}$ to give the correct $\eta_c$ mass. (This small shift in the overall
energy normalization can be eliminated by a slight change of $\sim 50{\rm MeV}$
in the quark masses used by Eichten et al.\cite{cornell} and a corresponding
small adjustment of the other parameters. To have a more direct comparison of
the wavefunctions and spectra, we refrained from making these adjustments
here.)

In Figs. \ref{fig2} and \ref{fig3} we show the eigenfunctions for the various
components of our quarkonium states (\ref{meson}) as well as the corresponding
eigenfunction of Eichten et al.\cite{cornell}.  The singlet component of our
ground state wave functions are very similar to those of the funnel potential.
At small radii this is true also for the excited states, since the coupling to
``octet'' components is proportional to $r$ in (\ref{diffeq}). The
``octet''-components increase with increasing excitation energy of the
quarkonium and lead to the appearance of extra nodes in the higher lying
``singlet'' wave functions  at large radii (since this is a coupled channel
problem, the extra nodes do not mean that we missed some bound states). As
noted earlier, the $E$-components of the meson state contribute negatively to
its norm.  All the stable quarkonia states we found are however positive norm
states.  We could not obtain any stable state where the octet components are
dominant.

Let us speculate at this point on the fact that only very few stable quarkonium
states were found. This is of course quite in line with the experimental
observation that only a few heavy quarkonia are stable against decay by strong
interactions. For reasons which we had not anticipated, this basic property
seems already to be incorporated in the nonhermitian coupling to
$E$-components.

The strength of this coupling in our model is however determined by the gluon
condensate $\phi^2$, which is not expected to vanish even in the purely gluonic
theory.  From table \ref{tab3} we see that the instability sets in at an
excitation
energy of between $1-1.3{\rm GeV}$ in this model. Since we cannot account for
the  decay into light $q\bar q$-mesons with a parameter which is essentially
independent of the number of (light) flavours, we speculate that the
nonhermitian coupling proportional to $\phi$ effectively accounts for the
possible decay channel
\begin{equation}
{\rm Quarkonium}^*\longrightarrow {\rm Quarkonium}+{\rm Gluonium}\,,
\end{equation}
in this model. From the fact that we do not seem to find any stable quarkonium
states more than $1.3{\rm GeV}$ above the ground state, we would estimate this
to be the threshold for the production of the lightest gluonium. This estimate
of the lightest gluonium mass $m_G \agt 1.3{\rm GeV}$, is in almost
perfect agreement with our previous interpretation of the energy shift
$-C=m_g\sim m_G/2$.

Since the production of light $q\bar q$-pairs has a much lower threshold, it is
this process which limits the stability of physical quarkonia.  We therefore
expect this model, which does not (not even effectively) incorporate this decay
channel, to still predict more stable states than are experimentally observed.

\section{Conclusion}
\label{concl}

We developed a hamiltonian formalism, which enabled us to estimate the effect
of a nonperturbative gluonic ground state on quarkonia in a systematic short
distance, weak coupling and $1/m$ expansion of the effective hamiltonian. The
gauge invariant basis\cite{basis} was extended to include color octet
quark-antiquark pairs coupled to vacuum fluctuations.  Hamiltonian matrix
elements in this basis are gauge invariant to the order in the short distance
expansion we considered.

After separating hard and soft gluons in the gauge (\ref{separ}), we obtained
the effective hamiltonian neglecting radiative corrections to the coulomb
interaction from hard gluons. We thus neglected the logarithmic corrections to
the effective coupling strength at very short distances. The correct behaviour
of the potential for $r<0.2 fm$ can in principle be included by ``hand'' in a
modification of the coulomb part of the interaction\cite{buch}.  Although the
quarkonium spectrum is not very sensitive to this correction at small
distances, it could become important for the evaluation of decay widths, which
depend on the wave function at the origin.

In the limit of very heavy quark masses, where all terms of order $1/m$ can be
neglected, an energy independent effective potential was obtained, which shows
an approximately linear behaviour at intermediate distances $0.4fm<r<0.8fm$
with an effective string tension of $\sim 840MeV$. In our approach this
behaviour of the potential arises due to the nonperturbative structure of the
gluonic vacuum parametrized by its gluon condensate and was not assumed from
the outset as in most phenomenological quark models.  This potential compares
favourably with recent lattice results\cite{schil}, the discrepancies at very
small $r<0.2fm$ being due to our neglect of radiative corrections.
Surprisingly, our potential is almost exactly reproduced  by an instantaneous
interaction derived from the effective gluon exchange proposed by
Gribov\cite{gribov}. Our short distance expansion for the effective potential
however is only valid for $r<0.9fm$, beyond which the potential aquires an
imaginary part. An extension of the model to larger distances would require a
more detailed knowledge of the vacuum structure in the form of higher
dimensional condensates, or some other effective parametrization of this
structure. The approach in this case would become increasingly phenomenological
and also more complicated in this case. Its predictive power is therefore
probably limited to heavy quarkonia, where a detailed knowledge of the
potential for very large radii does not seem necessary.

We showed that the numerical solution of the coupled channel problem for
vector- and pseudoscalar- quarkonia (they are degenerate to order $1/m$) is
feasible and an exact diagonalization of the hamiltonian in the extended basis
therefore possible. The resulting exact spectrum does not show the far too
rapid increase of the eigenvalues with the principal quantum number of the
perturbative approach to the vacuum effects\cite{volos}\cite{leut}. In addition
to the usual ``singlet'' wave-functions describing the quark and anti-quark of
the quarkonium when they are coupled in a colour singlet, we also obtain the
``octet'' components of quarkonium states describing the quarks in the octet
configuration when an additional (soft) gluon is around. Since hadronic decays
mainly proceed from this ``octet'' configuration with the creation of an
additional octett $q\bar q$-pair from a hard gluon, this approach opens the
possibility of estimating nonperturbative contributions to hadronic decays.

     We compare our results for the spectrum and wavefunctions with those of
the Cornell potential\cite{cornell} for pseudoscalar quarkonia in order $1/m$.
With the standard value for the gluon condensate and quark masses and coupling
constant used by the Cornell group we obtain the correct splitting between
$\eta_c$ and $\eta_c'$ and make predictions for the $\eta_b$'s. Our main
concern was however a better theoretical understanding and justification of the
phenomenological ingredients common to most nonrelativistic models for heavy
quarkonia and we refrained from adjusting the few parameters of this approach
to optimally reproduce the experimental data. A better description of the
potential at short distances with the inclusion of radiative corrections to the
coulomb force and the consideration of hadronic decay channels is clearly
desireable before a detailed comparison with phenomenology is attempted.

\section*{Acknowledgements}

	 We thank K.~Langfeld for help in the numerical solution of the coupled
equations and for useful discussions. One of the authors (C.A.A.N.) is grateful
for the kind hospitality extended to him at the Institut f\"ur Theoretische
Physik der Technischen Universit\"at M\"unchen. He would also like to thank
Professors H.~Leutwyler and K.~Yazaki for helpful discussions.  M.S. would like
to thank St.G\l azek for suggesting the extension of the basis. C.A.A.N. is
supported by CNPq/Brazil. M.S. is supported by Deutsche Forschungsgemeinschaft
with grant Scha/1-2.

\begin{figure}
\caption{
(a) Effective potential (solid curve) from (\protect\ref{pot})
and Coulomb potential (dotted curve)
with $\alpha_s=0.39$, $C=-756 \text{ MeV}$ and $\phi^2=(360
\text{ MeV})^4$. \protect\\
(b) Effective potential (solid curve) with the same parameters as in (a).
The potential extracted from lattice data\protect\cite{schil} with
$\protect\sqrt{\sigma}=365 \text{ MeV}$ (dot-dashed curve) and $\protect
\sqrt{\sigma}=505 \text{ MeV}$ (dashed curve). Gribov potential
\protect\cite{gribov} with
$\kappa=500 \text{ MeV}$
(crosses) and Cornell potential\protect\cite{cornell} (dotted curve) with
$\alpha_s=0.39$ and $a=2.34 \text{ GeV}^{-1}$.}
\label{fig1}
\end{figure}

\begin{figure}
\caption{
The wavefunctions of the ground, 1$^{\text{st}}$- and 2$^{\text{nd}}$-
excited pseudoscalar states of charmonium are shown in figures (a), (b) and
(c) respectively. The dashed curve is the wavefunction for the funnel
potential\protect\cite{cornell} for comparison. The solid curve is the singlet
component $S(r)$ of the quarkonium state in our calculation. The dot-dashed
and dotted curves are the ``octet'' components $E_1(r)$ and $E_2(r)$ of
equation (\protect\ref{diffeq}) respectively. The solution was obtained
with the
parameters $m_c=1840 \text{ MeV}$, $\alpha_s=0.39$, $\phi^2=(360
\text{ MeV})^4$ and
$C=-756 \text{ MeV}$.}
\label{fig2}
\end{figure}

\begin{figure}
\caption{
The wavefunctions of the ground, 1$^{\text{st}}$-, 2$^{\text{nd}}$- and 3$^{
\text{nd}}$ excited pseudoscalar states of bottonium are shown in figures (a),
(b), (c) and (d), respectively. The dashed curve is the wavefunction for the
funnel potential\protect\cite{cornell} for comparison. The solid curve is the
singlet
component $S(r)$ of the quarkonium state in our calculation. The dot-dashed
and dotted curves are the ``octet'' components $E_1(r)$ and $E_2(r)$ of
equation (\protect\ref{diffeq}) respectively. The solution was obtained with
the parameters $m_b=5170 \text{ MeV}$, $\alpha_s=0.39$, $\phi^2=(360
\text{ MeV})^4$ and $C=-756 \text{ MeV}$.}
\label{fig3}
\end{figure}

\begin{table}
\caption{The Coulomb energy in MeV is presented in the first column for the
ground state and first excitation of $c\bar c$ and $b\bar b$ (pseudoscalar or
vector). Second and third columns contain the nonperturbative contributions in
MeV calculated respectively within perturbation theory and with our method. We
used $m_c=1840 \text{ MeV}$, $m_b=5170 \text{ MeV}$, $\alpha_s=0.39$,
$\phi^2=0.012 \text{ GeV}^4$ and $C=-756 \text{ MeV}$.}
\label{tab1}
\begin{tabular}{cdcc}
& Coulomb & Pert. theory & Non-pert. \\
\hline
$\eta_c$, $J/\psi$ & -124.4 & 311 & 66 \\ $\eta_c'$, $\psi '$ & -31.1  & 21525
& 424 \\ $\eta_b$, $\Upsilon$ & -349.5 & 14 & 8\\ $\eta_b'$, $\Upsilon '$ &
-87.4  & 970 & 111 \\
\end{tabular}
\end{table}

\begin{table}
\caption{The Coulomb energy in MeV is presented in the first column for the
ground state and first excitation of  $b\bar b$ (pseudoscalar or vector) at two
different scales $\mu$. Second and third columns contain the nonperturbative
contributions in MeV calculated respectively within perturbation theory and
with our method. We used $C=-756 \text{ MeV}$, $\phi^2=0.042 \text{ GeV}^4$
and for $\mu=1.44 \text{ GeV}$: $m_b=4866 \text{ MeV}$, $\tilde
\alpha_s=0.38$. For $\mu=0.99 \text{ GeV} $: $m_b=5010 \text{ MeV}$,
$\tilde \alpha_s=0.54$.}
\label{tab2}
\begin{tabular}{cccc}
& Coulomb & Pert. theory & Non-pert. \\
\hline
$\eta_b$, $\Upsilon$ ($\mu=1.44 \text{ GeV}$) & -312 & 25 & 11 \\
$\eta_b'$, $\Upsilon'$ ($\mu=1.44 \text{ GeV}$) & -78  & 1762 & 129 \\
$\eta_b$, $\Upsilon$ ($\mu=0.99 \text{ GeV}$) & -649 & 6 & 3\\ $\eta_b'$,
$\Upsilon '$ ($\mu=0.99 \text{ GeV}$) & -162  & 396 & 72 \\
\end{tabular}
\end{table}

\begin{table}
\caption{Masses in MeV of pseudoscalar quarkonia with the funnel potential and
with our effective nonperturbative hamiltonian. Parameters used: $m_c=1840
\text{ MeV}$, $m_b=5170 \text{ MeV}$, $\alpha_s=0.39$, $\phi^2=(360 \text{ MeV}
)^4$, $C=-756 \text{ MeV}$.}
\label{tab3}
\begin{tabular}{ccc}
& funnel & nonpert. \\
\hline
$\eta_c$ & 2980 & 2980 \\ $\eta_c'$ & 3571 & 3594 \\ $\eta_c''$ & 3994 & 3993
\\ $\eta_b$ & 9213 & 9344 \\ $\eta_b'$ & 9805 & 9739 \\ $\eta_b''$ & 10150 &
10084 \\ $\eta_b'''$ & 10427 & 10610 \\
\end{tabular}
\end{table}

\end{document}